\documentclass[aps,prb,preprint,superscriptaddress,showpacs,showkeys]{revtex4}
\usepackage{amsfonts,amsmath,amssymb,bm}

\begin{document}

\title{Two-particle irreducible effective action approach to nonlinear current conserving approximations in driven systems}

\author{J. Peralta-Ramos}
\email{jperalta@df.uba.ar}

\affiliation{CONICET and Departamento de F\'isica, Facultad de Ciencias Exactas y Naturales, Universidad de Buenos Aires-Ciudad Universitaria, Pabell\'on I, 1428 Buenos Aires, Argentina}

\author{E. Calzetta}
\email{calzetta@df.uba.ar}
\affiliation{CONICET and Departamento de F\'isica, Facultad de Ciencias Exactas y Naturales, Universidad de Buenos Aires-Ciudad Universitaria, Pabell\'on I, 1428 Buenos Aires, Argentina}

\date{\today}

\begin{abstract}
Using closed-time path two-particle irreducible coarse-grained effective action (CTP 2PI CGEA) techniques, we study the response of an open interacting electronic system to time-dependent external electromagnetic fields. We show that the CTP 2PI CGEA is invariant under a simultaneous gauge transformation of the external field and the full Schwinger-Keldysh propagator, and that this property holds even when the loop expansion of the CTP 2PI CGEA is truncated at arbitrary order. The effective action approach provides a systematic way of calculating the propagator and response functions of the system, via the Schwinger-Dyson equation and the Bethe-Salpeter equations, respectively. We show that, due to the invariance of the CTP 2PI CGEA under external gauge transformations, the response functions calculated from it satisfy the Ward-Takahashi hierarchy, thus warranting the conservation of the electronic current beyond the expectation value level. We also clarify the connection between nonlinear response theory and the WT hierarchy, and discuss an example of an {\it ad hoc} approximation that violate it. These findings may be useful in the study of current fluctuations in correlated electronic pumping devices. 
\end{abstract}


\maketitle

\section{Introduction}
In this paper, we are interested in the response of an open electronic system to time-dependent external electromagnetic fields. Although our analysis is quite general and can be applied to a variety of condensed matter systems, we have in mind a system of {\it strongly} interacting electrons subjected to external driving fields and in contact with two ideal reservoirs of noninteracting electrons. This could be a picture of the so-called interacting electron pumping devices, which are currently attracting much interest both experimentally and theoretically\cite{pump}. In these devices, a direct current can be generated by applying slowly-oscillating external fields to the central electrons, even in the absence of a bias voltage difference between reservoirs. These are very interesting systems to study from the theoretical point of view, since they represent a unique and challenging combination of strongly correlated particles and quantum transport, and for which the study of current conservation beyond the expectation value level is nontrivial. A typical situation where one needs to go beyond the expectation value of the current is when dealing with current fluctuations. It is with these applications in mind that the following considerations were developed. 
 
The main purpose of this work is to determine the basic requirements that a field-theoretical approach to {\it open driven} systems must satisfy in order to produce current conserving results (in the sense of the WT hierarchy developed in Sec. \ref{wtintro}), in transport calculations {\it going beyond linear response}. Another aim we have in mind is to clarify the close relation that exists between current conservation and response theory, especially in the nonlinear regime. In order to analyze these issues, we combine the so-called external gauge invariance method with the closed-time path 2PI coarse-grained effective action (CTP 2PI CGEA), suitable for the description of strongly interacting quantum open systems both in and out of equilibrium\cite{calz88,libro,calzlan,calhu,calgauge}. 

The area of nonequilibrium physics in interacting systems is gaining increasing interest nowadays, particularly time-dependent quantum transport in correlated systems\cite{dah,bon,vel,myo,kr}. The so-called {\it time propagation method}\cite{dah} constitues a significant advance in this area. It  consists in first determining the (interacting) equilibrium Green function and then propagating it by using the Kadanoff-Baym equations. This is equivalent\cite{dah,bon} to solving the Bethe-Salpeter equation for the particle-hole propagator (which is related to the first order response function of the system), but numerically much less expensive. This is a powerful method in which external fields are treated exactly to all orders while many-body interactions are treated perturbatively. In this paper, we adopt another method to study nonequilibrium transport through correlated systems, in which both the external fields and the many-body interactions are treated perturbatively. Since it is based on an expansion of the Green function in powers of the external fields, it is only valid for weak external fields. On the other hand, it allows us to focus on current conservation beyond the expectation value level (i.e. conservation of current fluctuations). In this aspect the present work goes beyond previous analysis such as those described in Refs. \onlinecite{dah,bon,vel,myo,kr}.

The paper is organized as follows. In Sec. \ref{wtintro} we present the Ward-Takahashi (WT) hierarchy, which is the most general form of current conservation beyond the expectation value level, for a system driven by external fields. We discuss two possible ways (based on effective action techniques) of generating approximations to the non-equilibrium many-body problem that satisfy the WT hierarchy. One of such methods, relying on the external gauge invariance of the effective action, is used throughout the paper.  
In Sec. \ref{ctpea} we introduce the 2PI CGEA of the system, closely following Ref. \onlinecite{libro}. In Sec. \ref{egisec}  
we prove that the exact and truncated effective actions are external gauge invariant. 
Using these theoretical tools, we show in Sec. \ref{wtegisec} that for any approximation to the Schwinger-Dyson equation (obtained from a truncation of the loop expansion of the 2PI CGEA) there exists a corresponding approximation to the Bethe-Salpeter equations (which give the vertex functions), such that the WT hierarchy holds. The WT identities are systematically obtained from the external gauge symmetry of the 2PI CGEA. In this section we also clarify the relation between the WT hierarchy and current-conserving nonlinear response theory. In Sec. \ref{approxsec} we show, using a simple example, how {\it ad hoc} approximations to the 2PI CGEA (not resulting from a truncation of its loop expansion) violate the WT hierarchy. A brief summary is given in Sec. \ref{concsec}.

\section{Ward-Takahashi hierarchy and current conservation}
\label{wtintro}

In many-particle systems and particularly in quantum transport theories, $n$-point vertex functions play a fundamental role, since they represent generalized currents which satisfy the hierarchy of Ward-Takahashi identities\cite{rivier}. This hierarchy is satisfied to all orders in the exact theory, thus guaranteeing local gauge invariance and the conservation of the associated charges. 

There are two equivalent ways of obtaining the WT identities. The first one, due to Rivier and Pelka \cite{rivier} (See also Ref. \onlinecite{taka2}), relies on the equation of motion of the $n$-particle Green's function (GF). In the second one, due to Kadanoff and Baym \cite{baym,kb}, the two-point GF is expanded in powers of a fictitious non-local external field (for a recent development see Ref. \onlinecite{vel}). To zeroth order in the expansion, the classical continuity equation is recovered 
(i.e. the mean current is divergenceless); to first order in the external field, the usual WT identity relating the three-point vertex to the two-point propagator is obtained. Higher order terms in the expansion reproduce the hierarchy of WT identities obtained by the first method. We note that the second approach is not directly related to Baym's $\Phi-$derivable approximation\cite{baym,dedom,libro}, because in principle no approximation is involved. It is just an equivalent way of deriving the WT hierarchy which is based on a functional approach and not on the equations of motion for the propagators.

In its most general form, the generalized continuity equation (WT hierarchy) can be written as (we use the simplified notation $1=(t_1,\bf{r_1})$ and employ Schwinger-Keldysh non-equilibrium formalism\cite{kel,calz88,calzlan,calhu,chou,libro,rammer,kobes}): 
\begin{equation}
\begin{split}
& \partial_\mu^z \Lambda^\mu_{(n)}(1\ldots n,1'\ldots n';z) = \\
& i^n e \{ -\delta(z-n')G_n(1\ldots n,1'\ldots (n-1')z)+\ldots \\
& +(-1)^n \delta(z-1')G_n(1\ldots n,2'\ldots z)+\ldots \\
& +\delta(z-n)G_n(1\ldots (n-1)z,1'\ldots n')-\ldots \\
& -(-1)^n\delta(z-1)G_n(2\ldots z,1'\ldots n')\}
\end{split}
\label{jer}
\end{equation}
where 
\begin{equation}
\begin{split}
& \Lambda^\mu_{(n)}(12\ldots n,1'2'\ldots n';z) = \\
& <T_c j^\mu(z)\psi(1)\psi(2)\ldots\psi(n)\psi^\dagger(n')\ldots \psi^\dagger(2')\psi^\dagger(1')>
\end{split}
\label{lam}
\end{equation}
is the $(n+1)$-point vertex function with current insertion at $z=(t_z,\bf{r_z})$,  
\begin{equation}
\begin{split}
& j^\mu(z)=-e \lim_{z' \rightarrow z} D^\mu(z,z')\psi^\dagger(z')\psi(z) \qquad ;\\
& D^i(z,z') = (2i)^{-1}(\nabla^i_z-\nabla^i_{z'}) \qquad \qquad \mu=i=1,2,3 \qquad ,\\
& D^0(z,z') = 1 \qquad \qquad \qquad \qquad \qquad \qquad \mu=0
\end{split}
\label{curr}
\end{equation}
and $G_n$ are real-time propagators defined as usual 
\begin{equation}
\begin{split}
& G_{n}(1\ldots n,1'\ldots n') = \\
& i^{-(n)}<T_c \psi(1)\ldots \psi(n)\psi^\dagger(n')\ldots\psi^\dagger(1')> \qquad .
\end{split}
\label{gn}
\end{equation}
Note that $\Lambda_{(n=0)}^\mu(z) = <j^\mu(z)>$, and that 
\begin{equation}
D^i(z,z')[A] = -i\{\nabla^i_z-\nabla^i_{z'} - ie[A^i(z)+A^i(z')]\}/2 
\label{dop}
\end{equation}
in the presence of external fields. 

In Eqs. (\ref{lam}) and (\ref{gn}), $T_c$ is the closed-time path time ordering operator, $\psi$ and $\psi^\dagger$ are field operators in the Heisenberg representation, and $<\ldots>$ stands for an average taken with respect to the density matrix of the system in the remote past. The classical continuity equation and the usual WT identity\cite{ward} correspond to the cases $n$=0 and $n$=1 of Eq. (\ref{jer}), respectively:
\begin{equation}
\partial_\mu^z <j^\mu(z)> = 0 
\label{ejemplo0}
\end{equation}
and 
\begin{equation}
\partial_\mu^z \Lambda_{(1)}^\mu (1,1';z)= ie [-\delta(z,1')G(1,z)+\delta(z,1)G(z,1')] \qquad ,
\label{ejemplo1}
\end{equation}
being $G(1,1')$ the single-particle propagator. 

As already mentioned, in the exact theory Eq. (\ref{jer}) holds for arbitrary $n$. The hierarchy of $n$-particle propagators take into account that field excitations can be created and annihilated in the second-quantization formalism, thus acting as source terms for the generalized currents represented by the vertex functions. In other words, in the exact theory, particle number is {\it strongly} conserved, not only in the mean. This fact has an important consequence in quantum transport theories based on response theory to an external field, which is the main subject of this work. For a many-particle system in the presence of an external field, the hierarchy for the exact propagators implies that the current, as defined by Eq. (\ref{curr}), is conserved to all orders in the external perturbation, with linear response theory corresponding to the special case $n$=1. 

It is impossible, in general, to obtain the propagators of an interacting field theory exactly, and some approximations must be made. In the weak-coupling regime, no conflict arises between conservation laws and perturbative expansions based on Feynman diagrams, because conserved quantities are conserved order by order in the expansion\footnote{Of course, no conflict arises for non-interacting particles, since in that case the lifetime of field excitations is infinite. See Ref. \onlinecite{morg} for a discussion of second order response theory in non-interacting systems.}. On the other hand, for a strong coupled theory re-summation concepts are usually needed\cite{libro,dedom}, and warranting conservation laws then becomes a nontrivial issue.  
Symmetries satisfied at the exact level may not be satisfied by approximate propagators, thereby violating (generalized) current conservation dictated by the hierarchy in Eq. (\ref{jer}). 

A systematic way of generating conserving approximations (at the classical level, i.e. $n$=0) was given by Baym \cite{baym} and corresponds to his well-known $\Phi$-derivable scheme. The self-energy is obtained from a functional $\Phi$ consisting of an infinite series of two-particle irreducible (2PI) closed diagrams, constructed from full propagators and bare vertices\cite{dedom}. The solutions obtained from truncating the $\Phi$ functional are such that, if $\Phi$ is invariant under a simultaneous symmetry transformation of the classical field and propagator, the expectation values of the respective Noether currents are conserved\cite{baym,selfcons,libro}. We emphasize that this situation corresponds to the case $n$=0 of the WT hierarchy, given by Eq. (\ref{ejemplo0}). Therefore, the conservation of generalized currents encoded in the WT hierarchy is not automatically warranted in this approach. Moreover, in certain situations (e.g. in theories exhibiting spontaneous symmetry breaking) the hierarchy is already violated at the level of the self-energy (corresponding to $n=$1) which does not satisfy the Nambu-Goldstone theorem\cite{libro,hees,rei}. The reason is simple: the $n$-point functions obtained by functional differentiation in a $\Phi$-derivable approximation may not be equal to the one-particle irreducible\cite{libro} (1PI) functions that satisfy the WT hierarchy. 

To our knowledge, there are two possible approaches to overcome this problem present in $\Phi$-derivable approaches\footnote{We refer to approaches to solve the problem that still belong to the two-particle irreducible class. Of course, one could in principle solve the problem of local symmetries (WT identities) by treating the dynamics of higher order correlations. This would require {\it n}-particle irreducible effective actions, with $n > 2$ (see  Refs. \onlinecite{calzlan} and \onlinecite{bergesnpi}). However, we think that the relation between nonlinear response and current conservation is clearly appreciated in the 2PI formalism, which is also technically feasible as compared to $n > 2$ effective actions. It remains to be investigated the possibility of relating higher order response functions (and the response of higher order correlations) to higher order effective actions in the {\it n}-particle hierarchy.}. In the one put forward by van Hees and Knoll\cite{hees}, a non-perturbative approximation for the 1PI quantum effective action is obtained on top of a self-consistent solution to Schwinger-Dyson equations (SDE) derived from the truncated 2PI effective action. The vertex functions obtained in the usual way from this 1PI effective action fulfill the WT identities. The extra terms not accounted for in the $\Phi$-derivable approximation and required to satisfy the identities are encoded in a Bethe-Salpeter equation (BSE) and higher vertex equations (see Ref. \onlinecite{carr07} for a calculation of conductivity in QED). The second approach\cite{bando,wang} is based on the concept of {\it external gauge invariance}, and provides a systematic way of generating consistent SDE and BSEs that automatically satisfy WT identities. Most importantly in the context of transport theories of strongly interacting systems, the derivation of the SDE and the BSEs can be done, in principle, to any order in the external field coupled to the system. Therefore, the latter is especially suited for the study of response theory beyond first order and its relation to current conservation beyond the expectation value level, and will be employed here. 

\section{CTP 2PI coarse-grained effective action}
\label{ctpea}
In this section, we introduce the closed-time path 2PI effective action\cite{libro,calzlan,calhu,calgauge,calz88,rammer} for a specific fermionic system open to an environment. A detailed presentation of effective action techiques and the CTP formalism can be found in Ref. \onlinecite{libro}.
 
The system we are interested in consists of a central region with interacting electrons described by fields $(\psi^\dagger,\psi)$ and coupled to an external field. This central region is connected to two ideal reservoirs of noninteracting electrons described by fields $(\phi_\alpha^\dagger,\phi_\alpha)$, with $\alpha=(L,R)$ denoting left or right reservoir. The reservoirs are assumed to remain in equilibrium at all times. 
 
The CTP classical action of the system is $S[\bar\psi,\psi,\bar\phi,\phi] = S_\psi + S_\phi + S_c$, with:
\begin{equation}
\begin{split}
S_\psi &= c_{AB}\bar\psi^A C^{-1}_{AB} \psi^B + S_{int}[\bar\psi,\psi] \\
S_\phi &= c_{AB}\bar\phi^A B^{-1}_{AB} \phi^B \\
S_c &= c_{AB} \bar\psi^A T_{AB} \phi^B + \textrm{h. c.}
\end{split}
\label{claction}
\end{equation}
where 
\begin{equation}
S_{int}[\bar\psi,\psi] = \frac{1}{24} U_{ABCD} \bar\psi^A \psi^B \bar\psi^C \psi^D 
\label{sint}
\end{equation}
being $U_{ABCD}$ the completely antisymmetrized bare interaction local vertex. For the moment, we shall not specify the structure of the four-fermion vertex $U_{ABCD}$ since its precise form will not be needed until Sec. \ref{approxsec}. $T_{AB}$ is a local coupling parameter between the central region and the reservoirs. 

We are using a DeWitt notation\cite{libro,dewitt} with $A=(x,a)$, $x=(t_x,r_x,\sigma)$ and $a=(1,2)$ [or $(+,-)$] being CTP indices indicating the branch within the closed-time contour. For the fields describing electrons inside the reservoirs, an additional index $\alpha$ must be included in the CTP indices $(A,B)$, but for simplicity we leave it implicit. Repeated indices are assumed to be integrated or summed. $(\bar\psi,\psi)$ and $(\bar\phi,\phi)$ are Grassmann variables and $c_{ab}$ is a CTP metric $c_{ab}=\textrm{diag}(1,-1)$, while $c_{AB} = c_{ab}\delta(x_A,x_B)$. $C_{AB}$ and $B_{AB}$ are the free CTP propagators corresponding to $S_\psi$ and $S_\phi$. They satisfy (we set $\hbar,m,c=1$ in what follows)
\begin{equation}
\begin{split}
c_{AB} \Box B_{BC} &= i \delta_A^C \\
c_{AB} \tilde{\Box} C_{BC} &= i \delta_A^C
\end{split}
\label{free}
\end{equation}
where $\Box (1)= i\partial_{t_1} - h_0(1)$ while $\tilde{\Box}(1) = i\partial_{t_1} - \tilde{h}_0(1)$, being $h_0$ and $\tilde{h}_0$ single-particle Hamiltonians corresponding to the reservoirs and the central region, respectively. Explicitly, they read
\begin{equation}
\begin{split}
h_0(1)&= -\frac{1}{2}\nabla^2_1 \qquad \textrm{and} \\
\tilde{h}_0(1) &= \frac{1}{2}[-i\nabla_1 + e A_i(1)]^2 + eA_0(1) 
\end{split}
\end{equation}
where $A_\mu(z)$ is the external classical field and $i=(1,2,3)$. Note that it is not necessary to add CTP indices to the external field, since this is a physical source, and in any case we would obtain $A_{\mu,1}=A_{\mu,2}$. It is important to remark that, since $A_\mu$ is external, it is not treated as a dynamical field.

From the classical action of Eq. (\ref{claction}), we form the CTP generating functional $Z_{CTP}$ with local and bi-local external sources $J_A$ and $K_{AB}$, respectively. Since we are interested in the dynamics inside the central region coupled to reservoirs, only the fields $(\bar\psi,\psi)$ describing the ``system'' will be coupled to the external sources. $Z_{CTP}$ contains all the information about the non-equilibrium many-body system since all closed-time path correlators can be calculated from it\cite{libro}. When written as a path integral over field configurations it reads
\begin{widetext}
\begin{equation}
Z_{CTP}[J,K] = \int [D\bar\phi][D\phi][D\bar\psi][D\psi] 
\exp{i(S[\bar\psi,\psi,\bar\phi,\phi] + \bar{J}_A\psi^A +\bar \psi^A J_A + \frac{1}{2} K_{AB}\bar\psi^A \psi^B)}
\end{equation}
\end{widetext}
where $(\bar J_A,J_A)$ and $K_{AB}$ are fermionic and bosonic sources, respectively. The measure corresponding to each field, for example $\psi$,  actually stands for $[D\psi_a]$ with $a=(1,2)$ the branch index. The CTP boundary condition of the path integral giving $Z_{CTP}$, namely the continuity of field histories in the remote future, is implicit. We assume that the initial state is prepared in the remote past, corresponding to the {\it in-}vacuum\cite{libro,calz88}. 
Note that $(\bar{J}_A,J_A)$ act as Lagrange multipliers constraining the deviations of $(\bar\psi,\psi)$ from their mean values, while $K_{AB}$ constrains their fluctuations\cite{calhu}. This is the main idea behind $n$PI effective action techniques, namely, to treat $1\ldots n-$point correlators on the same footing\cite{libro,bergesnpi,calhu}. By using the 2PI CGEA, only the one and two-point correlators are treated as dynamical variables and therefore obtained variationally; higher correlations (with $n>2$) are calculated from them. Note that the use of the 2PI CGEA to obtain equations of motion for mean fields and propagators represents an enormous simplification of the usual perturbation expansion. Moreover, nonperturbative approximation schemes based on expansions of the 2PI EA provide powerful calculational tools where standard expansion schemes break down\cite{libro,calgauge}.

Any correlation function of the many-body system can be obtained from $Z_{CTP}$ by functional differentiation with respect to $(\bar{J}_A,J_A)$ and then setting all sources to zero. For example, the CTP two-point propagator of the central region is 
\begin{equation}
\begin{split}
&<T_c \psi^A(\psi^B)^\dagger> = \frac{\delta^2 Z_{CTP}[J,K]}{i\delta_R \bar{J}_A i\delta_L J_B}\Big|_{[\bar J,J,K] = 0} =\\
& \int [D\bar\phi][D\phi][D\bar\psi][D\psi] \psi^A \bar{\psi}^B 
\exp{i(S[\bar\psi,\psi,\bar\phi,\phi])} \qquad ,
\end{split}
\label{gfun}
\end{equation}
with $\delta_{R,L}$ denoting right and left differentiation, respectively. Note that there is no need to use contour-ordering operators in the second line of Eq. (\ref{gfun}), since the path integral automatically arranges operators in the correct order. 

Since the reservoirs' fields enter the action quadratically, we can integrate them out exactly in $Z_{CTP}$, thus defining a new generating functional for CTP propagators belonging to the system. This {\it coarse-grained} generating functional reads: 
\begin{equation}
\tilde{Z}_{CTP}[J,K] = \int [D\bar\psi][D\psi] \exp{i(\tilde{S}[\bar\psi,\psi] + \textrm{sources})}
\label{ztilde}
\end{equation}
where the effective classical action is given by 
\begin{equation}
\tilde{S}[\bar\psi,\psi] = \tilde{S}_\psi = c_{AB}\bar\psi^A D^{-1}_{AB} \psi^B + S_{int}[\bar\psi,\psi]
\label{effclass}
\end{equation}
with
\begin{equation}
D^{-1}_{AB} = C^{-1}_{AB} + i\Sigma_{\phi,AB} \qquad .
\label{d}
\end{equation}
The self-energy $\Sigma_{\phi}$ describes the influence of the reservoirs on the particles' dynamics inside the system, and it is given explicitly by 
\begin{equation}
\Sigma_{\phi,AD} = T_{AB}^* B_{BC} T_{CD}
\label{sigmaphi}
\end{equation}
where $B$ is the given by Eq. (\ref{free}). It is a complex quantity whose imaginary part represents the tunneling rate of particles from the central region to the reservoirs. Note that $\Sigma_\phi$ can be easily calculated since it depends on the equilibrium CTP propagators of the reservoir composed of non-interacting electrons. The CTP propagators corresponding to the system are calculated from $\tilde{Z}_{CTP}$ using the analog of Eq. (\ref{gfun}) with the replacement $S \rightarrow \tilde{S}$. 

From $\tilde{Z}_{CTP}$ we define the CTP generating functional of connected propagators $W_{CTP}$ in the usual way, i.e. $\tilde{Z}_{CTP}=\exp{iW}$, and then apply a double Legendre transform to obtain the 2PI-CTP effective action of the system. The sources $(J,K)$ are connected to the mean field and full propagators (denoted by $G$) through 
\begin{equation}
\begin{split}
\frac{\delta W}{\delta \bar J_A} &= \hat{\psi}^A = 0 \\
\frac{\delta W}{\delta J_A} &= \hat{\bar \psi}^A = 0 \\
\frac{\delta W}{\delta K_{AB}} &= \frac{1}{2}G_{AB}
\end{split}
\label{deltaw}
\end{equation}
where we have used that the fermionic mean-fields $(\hat\psi,\hat{\bar\psi})=0$ since no symmetry breaking occurs. By performing the double Legendre transform on $W$ and then using the background field method we can write the CTP 2PI CGEA of the system as\cite{libro} (Tr and ln operations are understood in a functional sense)
\begin{equation}
\Gamma[G,A] = i \textrm{Tr ln} G - i D^{-1}_{AB} G^{AB} +\Gamma_2[G]
\label{gamma}
\end{equation}
where $A$ is the external classical field introduced before. We note that the CTP 2PI CGEA depends explicitly on the external field only through  $D^{-1}[A]$; this will be important in what follows. 

In Eq. (\ref{gamma}), $\Gamma_2[G]$ encodes all quantum corrections and consists of vacuum 2PI closed diagrams with full propagators in internal lines and vertices corresponding to a theory with shifted classical action $S[\hat{\bar\psi}+\bar\psi,\hat\psi+\psi]$ (neglecting constant and linear terms), where $(\bar\psi,\psi)$ denote fluctuations. The diagrams are vacuum because the mean value of the fluctuation fields is zero by construction. 
Note that because of the vanishing of the mean fields, the shifted classical action vanishes when evaluated at $(\hat{\bar\psi},\hat\psi)$. Therefore, the vertices contributing to $\Gamma_2$ are identical to those of the original classical action, namely, the quartic one with $U$ as coupling parameter. The case of noninteracting electrons corresponds in this scheme to $\Gamma_2=0$.
  
The Schwinger-Dyson equations for the CTP propagators follow directly from $\Gamma[G,A]$
\begin{equation}
\frac{\delta \Gamma}{\delta G_{AB}} = -\frac{1}{2} K_{AB}
\end{equation}
where the physical case corresponds to vanishing external sources $K = 0$. Defining the self-energy 
\begin{equation}
\Sigma_{AB}=-2 \delta \Gamma_2/\delta G^{AB}
\end{equation}
we can rewrite the SDE in the usual way
\begin{equation}
G_{AB}^{-1} = D^{-1}_{AB} +i\Sigma_{AB} \qquad ,
\end{equation}
being $\Sigma_{AB}$ one-particle irreducible by construction. 
 
In this paper, we shall not go beyond second order in the interaction parameter $U$, corresponding in this theory to a three-loop expansion. To this order, $\Gamma_2$ reads in our compact notation: 
\begin{equation}
\begin{split}
\Gamma_2 =& -\frac{1}{8} U_{ABCD}G_{AB}G_{CD} \qquad + \\
& i\frac{1}{48} U_{ABCD}U_{EFGH}G_{AE}G_{BF}G_{CG}G_{DH} \qquad .
\end{split}
\label{gam2}
\end{equation}
The first term corresponds to the so-called {\it double-bubble} diagram and the second term to the {\it basketball} diagram. In the SDE, the first term yields the time-dependent Hartree-Fock approximation, while the second is non-local and complex (therefore including fluctuation damping)\cite{calz88,libro,calhu}. We will return to these approximations when proving the external gauge invariance of the CTP 2PI CGEA, in Sec. \ref{egisec}, and also in Sec. \ref{approxsec}. 

\section{External gauge invariance of the CTP 2PI EA}
\label{egisec}

The crucial observation that allows us to relate the CTP 2PI CGEA of the open system to nonlinear transport through it is that $\Gamma[G,A]$ is invariant under a gauge transformation of the external field $A_\mu$. Following Bando, Harada and Kugo (Ref. \onlinecite{bando}), we will call this {\it external gauge invariance} (EGI) of the 2PI EA. We will first give a brief proof of the EGI of the {\it exact} 2PI CGEA, where exact means that $\Gamma_2$ is expanded to all loop orders in Eq. (\ref{gamma}). Then, we will prove that even when the loop expansion of $\Gamma_2$ is truncated at certain order, the truncated 2PI EA still remains EGI. 
 
Under a local transformation $U(1)=\exp{ie\varphi(1)}$, the external field and the full propagator transform as\cite{wang,rei} (we omit the CTP indices for the moment)
\begin{equation}
\begin{split}
& A_\mu \rightarrow A_\mu' = U A_\mu U^{-1} - i(\partial_{\mu} U) U^{-1}  \\
& G(1,2) \rightarrow G'(1,2) = U(1)G(1,2)U^{-1}(2) \qquad .
\end{split}
\label{egi}
\end{equation}
It is rather straightforward to prove that, if we retain all terms in the loop expansion of $\Gamma_2$, then the 2PI CGEA is EGI\cite{libro,rei,calgauge}. This can be simply understood recalling that the exact 2PI CGEA is precisely the generating functional of 1PI propagators. 
The proof of the EGI of the exact 2PI CGEA is based on the fact that, under a gauge transformation of the fields $(A_\mu,\bar{\psi},\psi)$, the CTP generating functional $\tilde{Z}_{CTP}[J,K]$ as given in Eq. (\ref{ztilde}) is invariant. This is because the transformation is equivalent to a change of integration variables in the path integral, whose Jacobian is trivial due to the unitarity of the transformation (det$~U=1$). Since the classical (effective) action $\tilde{S}[\bar{\psi},\psi]$ is external gauge invariant, only the source terms are transformed. For a local infinitesimal variation of the fields $\zeta=(A_\mu,\bar{\psi},\psi)$, given in compact notation by 
\begin{equation}
\delta \zeta = e(0,-i\varphi \bar{\psi},i\varphi \psi) + (\partial_\mu \varphi,0,0)
\label{zetafield}
\end{equation}
we obtain from Eq. (\ref{ztilde})
\begin{equation}
\begin{split}
\delta \tilde{Z}_{CTP} &= 0 = <\bar{J}_A\delta \psi^A + \delta \bar{\psi}^A J_A + \\
& \frac{1}{2}K_{AB}\delta \bar{\psi}^A\psi^B + \frac{1}{2}K_{AB}\bar{\psi}^A \delta \psi^B> ~,
\end{split}
\label{deltaz}
\end{equation}
with  
\begin{equation}
<\ldots> = \int [D\bar{\psi}][D\psi] (\ldots) \exp{i(\tilde{S}[\bar{\psi},\psi]+\textrm{sources})} \qquad .
\end{equation}
Using that
\begin{equation}
\begin{split}
\frac{\delta \Gamma}{\delta \bar{\psi}^A} &= -J_A + \frac{1}{2} K_{AB} \psi^B \\
\frac{\delta \Gamma}{\delta \psi^A} &= \bar{J}_A - \frac{1}{2} K_{AB} \bar{\psi}^B \\
\frac{\delta \Gamma}{\delta G_{AB}} &= -\frac{1}{2} K_{AB} \qquad ,
\end{split}
\end{equation}
which follows since $W$ and $\Gamma$ are Legendre transforms of each other, we can turn Eq. (\ref{deltaz}) into an equation for $\Gamma$:
\begin{equation}
\sum_{\alpha=2,3} \delta \zeta_\alpha \frac{\delta \Gamma}{\delta \zeta^\alpha} + \frac{\delta \Gamma}{\delta G_{AB}} \delta G^{AB} = 0 
\label{zj}
\end{equation}
where the variation of the full propagator is given by
\begin{equation}
\delta G^{AB}=\delta <\psi^A\bar{\psi}^B> = <\delta \psi^A \bar{\psi}^B + \psi^A \delta \bar{\psi}^B> \qquad .
\label{deltag}
\end{equation}
According to Eq. (\ref{zetafield}), Eq. (\ref{deltag}) is the infinitesimal version of the transformation rule for $G$ given in Eq. (\ref{egi}). Therefore, Eq. (\ref{zj}) shows that the exact 2PI CGEA is gauge invariant under Eq. (\ref{egi}). 

We turn now to the invariance of the 2PI effective action which results from a truncation in the loop expansion of the quantum correction $\Gamma_2$. For completeness, we will also show explicitly that the first two terms in Eq. (\ref{gamma}), the one-loop terms, are EGI.  
We start with the first term of $\Gamma[G,A]$, which is clearly EGI:
\begin{equation}
\textrm{Tr ln} G \rightarrow \textrm{Tr ln} (UGU^{-1}) = 
\textrm{Tr}[U (\textrm{ln} G) U^{-1}] = \textrm{Tr ln} G .
\end{equation}
The EGI of the second term of Eq. (\ref{gamma}) is proved by noting that $D^{-1}$ transforms as $G$ under Eq. (\ref{egi}). This follows directly from the equation of motion for the system's free propagator $C$, given in Eq. (\ref{free}), and the fact that $\Sigma_\phi$ is invariant under the external gauge transformation. The latter is a consequence of the equation of motion satisfied by $B$, the reservoir's free propagator, and the dependence of $\Sigma_\phi$ on this propagator, given in Eq. (\ref{sigmaphi}). 

The third term contributing to $\Gamma[G,A]$ is $\Gamma_2[G]$, the sum of 2PI vacuum diagrams with $G$ in internal lines. The proof of external gauge invariance proceeds as before, and relies on the transformation law for propagators and the structure of the 2PI EA. To be explicit, we rewrite the first term of Eq. (\ref{gam2}) taking into account that the interaction vertex $U_{ABCD}$ is local. Using the locality of the bare vertex  
we obtain for the double-bubble diagram 
\begin{equation}
\Gamma_2^{(1)} = -\frac{\tilde{U}}{8}  G_{AB}G_{BA}
\label{tdhf}
\end{equation}
which shows that, as already mentioned, the self-energy derived from it is local. It is clear from the above expression that, under the external gauge transformation given in Eq. (\ref{egi}), $\Gamma_2^{(1)}$ is invariant. The structure of higher terms in the loop expansion of $\Gamma_2$ is such that the EGI holds to arbitrary order. 

The main conclusion of this discussion is that, even though the system is open to reservoirs, its 2PI coarse-grained effective action is, order by order in the loop expansion, invariant under the gauge transformation given in Eq. (\ref{egi}). 
As we shall see, this makes the combination of the CTP 2PI-CG effective action approach and the external gauge invariance a powerful technique to study strongly interacting driven systems beyond linear response. We should remark that in the presence of non-vanishing mean fields, the 2PI CGEA may not be EGI order by order in a loop expansion\cite{libro,rei,calgauge}. 

\section{WT hierarchy from external gauge invariance}
\label{wtegisec}

We have seen that the CTP 2PI CGEA describing the system is left invariant under an external gauge transformation, and that this property holds to all orders in the (loop) expansion of the quantum corrections given by $\Gamma_2$. This implies that the SDE derived functionally from the 2PI effective action is automatically external gauge covariant. Most importantly, this is true for an {\it arbitrary} external classical field. This will be the key property to relate this symmetry with nonlinear response, thus generating approximations which satisfy the WT hierarchy to arbitrary order. We will show that the external gauge covariance is also inherited by the BSE for vertex functions, which as a consequence of EGI will satisfy the Ward-Takahashi hierarchy.  

Gauge invariance of the effective action $\Gamma[G,A]$ implies 
\begin{equation}
\Gamma[G,A] = \Gamma[G',A']
\end{equation}
where a prime denotes external gauge transformated quantities according to Eq. (\ref{egi}). 
The SDE for the CTP propagators is given by $\delta\Gamma/\delta G_{AB} = 0$, and due to the EGI of the 2PI CGEA, it is (external) gauge covariant:
\begin{equation}
\frac{\delta \Gamma[G,A]}{\delta G_{AB}} = U^{-1} \frac{\delta \Gamma[G',A']}{\delta G'_{AB}} U \qquad .
\end{equation}

This property of the 2PI CGEA has an important consequence. It implies that the variational procedure by which one obtains the SDE from the 2PI CGEA is independent of the external gauge. That is to say, if $G_{AB}[A]$ is a solution of the SDE in the background classical field $A$, i.e.
\begin{equation}
\frac{\delta \Gamma[G,A]}{\delta G_{AB}}\Big|_{G_{AB}=G_{AB}[A]} = 0 \qquad ,
\end{equation}
then the solution corresponding to a transformed external field $A'=UAU^{-1}-i(\partial U)U^{-1}$ is precisely $G'=UGU^{-1}$:
\begin{equation}
\frac{\delta \Gamma[G',A']}{\delta G'_{AB}}\Big|_{G'_{AB}=UG_{AB}[A]U^{-1}} = 0 \qquad .
\end{equation}
More explicitly, EGI of the 2PI CGEA implies 
\begin{equation}
G[A']=UG[A]U^{-1} \qquad .
\end{equation}

We will now make a connection between EGI and nonlinear response. For notational simplicity, CTP indices are omitted in the following. The solution $G[A]$ to the SDE can be expanded in powers of the external field $A_\mu$ (see Refs. \onlinecite{chou,heinz,kobes}): 
\begin{widetext}
\begin{equation}
\begin{split}
G[A](X,Y) &= G[0](X,Y) + iA_\mu \Pi_3^\mu + \frac{i^2}{2} A_\mu A_\nu \Pi_4^{\mu\nu} + \frac{i^3}{3}A_\mu A_\nu A_\rho \Pi_5^{\mu \nu \rho} \ldots \\
&= \sum_{n=0} \frac{i^n}{n!} \int d1\ldots dn A_{\mu_1}(1)\ldots 
A_{\mu_n}(n) \Pi_{(n+2)}^{\mu_1\ldots \mu_n}(1,\ldots,n; X,Y) \qquad ,
\end{split}
\label{expans}
\end{equation}
\end{widetext}
where in the second line we have made explicit the internal (integration) $(1,\ldots,n)$ and the external $(X,Y)$ variables. The dummy indices $\mu,\nu,\rho$, etc. of the first line are denoted by $\mu_1,\ldots,\mu_n$ in the second. 

The ``response'' functions $\Pi_{(n+2)}$ encode the variation of the full propagator with the external field. We note that it is possible\cite{heinz,chou}, in principle, to expand higher order correlation functions similarly to the two-point function as given in Eq. (\ref{expans}), but in this paper we shall only deal with $G[A](X,Y)$. Eq. (\ref{expans}) is given in the so-called\cite{chou,heinz} single-time representation of CTP correlators, which is simpler for calculations than the ``physical'' or {\it r/a} representation. Both representations are related by a similarity transformation characterized by an orthogonal matrix\cite{kel,chou} $Q=(\hat{1}-i\sigma_2)/\sqrt{2}$ with 
\begin{equation}
\sigma_2 = \left( \begin{array}{cc}
0 &~ -i \\
i &~ 0 
\end{array} \right) ~ ; ~ \hat{1}=\left( \begin{array}{cc}
1 &~ 0 \\
0 &~ 1 
\end{array} \right) \qquad ,
\end{equation}
and are completely equivalent to each other: 
\begin{equation}
\begin{split}
& E_{i_1\ldots i_n}(1\ldots n)=2^{n/2-1}Q_{i_1\alpha_1}\ldots Q_{i_n\alpha_n} E_{\alpha_1\ldots \alpha_n}(1\ldots n) \\
& E_{\alpha_1\ldots \alpha_n}(1\ldots n) = 2^{1-n/2}Q^T_{\alpha_1 i_1}\ldots Q^T_{\alpha_n i_n} E_{i_1\ldots i_n}(1\ldots n)
\end{split}
\label{physra}
\end{equation}
where $E(1\ldots n)$ denotes a $n-$point CTP function, $\alpha_j$ ($i_j$) corresponds to the single-time (physical) representation and repeated indices are assumed to be summed. To give an example of Eq. (\ref{physra}), for the propagators in the physical 
\begin{equation}
\tilde{G} = 
\left( \begin{array}{cc}
0 & G_a \\
G_r & G_c 
\end{array} \right) 
\label{gphys}
\end{equation}
and in the single-time 
\begin{equation}
\tilde{G} = 
\left( \begin{array}{cc}
G_{++} & G_{+-} \\
G_{-+} & G_{--} 
\end{array} \right) 
\end{equation}
representations we have\cite{libro}:
\begin{equation}
\begin{split}
\tilde{G}_{11}&=\frac{1}{2}(G_{++}+G_{--}-G_{+-}-G_{-+}) = 0 \\ 
\tilde{G}_{12}&=G_a=\frac{1}{2}(G_{++}-G_{-+}+G_{+-}-G_{--}) \\
\tilde{G}_{21}&=G_r=\frac{1}{2}(G_{++}-G_{+-}+G_{-+}-G_{--}) \\
\tilde{G}_{22}&=G_c=\frac{1}{2}(G_{++}+G_{--}+G_{+-}+G_{-+}) \\
&= G_{+-}+G_{-+} = G_{++}+G_{--} \qquad .
\end{split}
\label{gtilde}
\end{equation}
The first equality in Eq. (\ref{gtilde}) is valid in general: any CTP function in the physical representation with all its indices set to ``1'' satisfies $\tilde{G}_{11\ldots1}=0$. The last line results from the identity $G_{+-}+G_{-+} = G_{++}+G_{--}$, which follows from the normalization of the step function $\theta(1,2)+\theta(2,1)=1$. 
In these expressions, $G_{r}$, $G_a$ and $G_c$ are retarded, advanced and correlation functions, respectively: 
\begin{equation}
\begin{split}
G_r(1,2) &= -i\theta(1,2)<\{\psi(1),\psi^\dagger(2)\}> \\
G_a(1,2) &= -i\theta(2,1)<\{\psi(1),\psi^\dagger(2)\}> \\
G_c(1,2) &= -i<\{\psi(1),\psi^\dagger(2)\}>
\end{split}
\end{equation}
where $\{,\}$ is the anticonmutator and $\theta$ the step function, while $G_{++,--,+-,-+}$ are the chronological (Feynman), antichronological (Dyson), lesser and greater correlations:
\begin{equation}
\begin{split}
G_{++}(1,2) &= -i<T \psi(1)\psi^\dagger(2)> \\
G_{+-}(1,2) &= i<\psi^\dagger(2)\psi(1)> \\
G_{-+}(1,2) &= -i<\psi(1)\psi^\dagger(2)>\\
G_{--}(1,2) &= -i<\tilde{T}\psi(1)\psi^\dagger(2)> \qquad ,
\end{split}
\end{equation}  
with $(T,\tilde{T})$ denoting the chronological and antichronological time ordering operator. 
Using this transformation rule, it is a simple matter to re-express Eq. (\ref{expans}) (and what follows from it) in the physical representation (see Refs. \onlinecite{chou} and \onlinecite{kobes}), but for clarity we will continue employing the single-time representation. 

We note that, as discussed in Ref. \onlinecite{chou}, in the physical representation the observables are given by retaining only the ``2'' component of the $n-$point function $\tilde{G}(1\ldots n)$, i.e. by the correlation functions $\tilde{G}_{2\ldots 2}$. For example, the two-point observable ($n=2$) is the fully symmetrized correlation function $\tilde{G}_{22}(X,Y)=G_c(X,Y)$, given in Eq. (\ref{gtilde}). Using the relation between the physical and single-time representations, the response of the correlation two-point function $\tilde{G}_{22}$ is immediately obtained
\begin{equation}
\begin{split}
\tilde{G}_{22}[A](X,Y) =& \tilde{G}_{22}[0](X,Y) ~+ \\
& i\tilde{\Pi}^\mu_{221}(X,Y;1)A_\mu(1) + \ldots 
\end{split}
\end{equation}
It is worth remarking that the use of the closed-time path method automatically ensures that the response functions are causal (see, for instance, Refs. \onlinecite{libro} and \onlinecite{chou})\footnote{The causality of the response functions is easily proved in the physical representation, and follows directly from the structure of the generating functional of CTP correlation functions.}. 

Returning now to Eq. (\ref{expans}), the response functions are given by
\begin{equation}
\begin{split}
& \Pi_{(n+2)}^{\mu_1\ldots \mu_n}(1,\ldots,n ; X,Y) = \\
& i^{-n}\frac{\delta G(X,Y;A)}{\delta A_{\mu_1}(1)\ldots \delta A_{\mu_n}(n)}
\Big|_{[A_{\mu_1} \ldots A_{\mu_n}] = 0}
\end{split}
\label{pidef}
\end{equation}
and correspond to $(n+2)-$point functions with $n$ current vertices inserted at locations $(1,\ldots,n)$ where interactions between the current and the external classical field take place. In more detail, the response functions for $n=1,2$ read
\begin{equation}
\begin{split}
\Pi_3^\mu(X,Y;z) &= -i<T_c j^\mu(z) \psi(X)\psi^\dagger(Y) > \qquad \textrm{and} \\
\Pi_4^{\mu\nu}(X,Y;z,w) &= -<T_c j^\mu(z) j^\nu(w) \psi(X)\psi^\dagger(Y) >
\end{split}
\label{pi12}
\end{equation}
Note that $-i\Pi_2(X,Y)$ corresponds to the propagator in the absence of the external field, which can be written as $G[0](X_i,Y_i;X_0-Y_0)$ due to time-translation invariance of the system in equilibrium. 

The structure of $\Pi_{(n+2)}$ as given in Eq. (\ref{pi12}) follows because functional differentiation with respect to $A_\mu$ yields a current insertion to which the external field couples. In functional language we have that 
\begin{widetext}
\begin{equation}
\begin{split}
& \frac{\delta G(X,Y;A)}{\delta A_{\mu_1}(1)\ldots \delta A_{\mu_n}(n)}
\Big|_{[A_{\mu_1} \ldots A_{\mu_n}] = 0} = \\
& \int [D\psi][d\bar{\psi}] \psi(X)\bar\psi(Y) \frac{\delta }{\delta A_{\mu_1}(1)\ldots \delta A_{\mu_n}(n)}
\Big|_{[A_{\mu_1} \ldots A_{\mu_n}] = 0} \exp{i(\tilde{S}[\bar\psi,\psi]+\textrm{sources})} \qquad .
\end{split}
\label{gader}
\end{equation}
\end{widetext}
Eq. (\ref{gader}) automatically leads to Eq. (\ref{pi12}) since 
\begin{equation}
\begin{split}
&\frac{\delta }{\delta A_{\mu_1}(1)\ldots \delta A_{\mu_n}(n)}
\Big|_{[A_{\mu_1} \ldots A_{\mu_n}] = 0} \exp{i(\tilde{S}[\bar\psi,\psi]+\textrm{sources})} \\
&= (-i)^n j_{\mu_1}(1)\ldots j_{\mu_n}(n) \exp{i(\tilde{S}[\bar\psi,\psi,A=0]+\textrm{sources})} \qquad .
\end{split}
\end{equation}

The functions $\Pi_{(n+2)}$ are obtained from the SDE by functional differentiation with respect to $A_\mu$ (and then setting $A=0$). This results in the BSEs for the response functions. For example, the BSE that determines $\Pi_3$ reads 
\begin{equation}
\frac{\delta G^{-1}}{\delta A_\mu} = \frac{\delta D^{-1}}{\delta A_\mu} + i \frac{\delta \Sigma}{\delta A_\mu} 
\end{equation}
where both sides of the equation are evaluated at $A_\mu = 0$. 
Using that $G(1,2)G^{-1}(2,1')=\delta(1,1')$ (where the coordinate $2$ is integrated), which implies 
\begin{equation}
\frac{\delta G^{-1}(1,2)}{\delta A_\mu(3)} = -\int d4 d5 G^{-1}(1,4) \frac{\delta G(4,5)}{\delta A_\mu(3)}G^{-1}(5,2)
\end{equation}
and defining 
\begin{equation}
\tilde{\Pi}_{(n+2)}^{\mu_1\ldots \mu_n} = G^{-1} \Pi_{(n+2)}^{\mu_1\ldots \mu_n} G^{-1} \qquad ,
\end{equation}
the BSE for $\Pi_{3}$ can be written as
\begin{equation}
\tilde{\Pi}_3^\mu = -\frac{\delta D^{-1}}{\delta A_\mu} - i \frac{\delta \Sigma}{\delta A_\mu} \qquad .
\label{pi3}
\end{equation}
Since the self-energy $\Sigma$ is defined through $\Gamma_2$, it is a functional of the full propagator $G$ and therefore the last term in Eq. (\ref{pi3}) involves $\Pi_3$. Namely, 
\begin{equation}
\begin{split}
\tilde{\Pi}_3^\mu(X,Y;z) &= -\frac{\delta D^{-1}}{\delta A_\mu} - i \frac{\delta \Sigma}{\delta G}\frac{\delta G}{\delta A_\mu} =\\ 
&-\frac{\delta D^{-1}(X,Y)}{\delta A_\mu(z)} - i \frac{\delta \Sigma(X,Y)}{\delta G(V,W)}\frac{\delta G(V,W)}{\delta A_\mu(z)} =\\
&-\frac{\delta D^{-1}(X,Y)}{\delta A_\mu(z)} - i \frac{\delta \Sigma(X,Y)}{\delta G(V,W)}\Pi_3^\mu(V,W;z) \qquad ,
\end{split}
\label{pi3det}
\end{equation}
where in the last two lines the coordinates are shown explicitly ($V$ and $W$ are integrated). This is an integral equation that determines $\Pi_3$ once the quantity $\delta \Sigma/\delta G$ is known. Taking further functional derivatives of the SDE with respect to $A_\mu$ (and setting $A=0$) results in a set of integral equations for $n>1$ response functions. Note that $\Sigma$ is directly obtained from $\Gamma[G,A]$, so the SD and BS equations are fully consistent with each other.

As we have seen, the combination of the SDE and the BSEs completely determine the full propagator and the response functions $\Pi_{(n+2)}$. The SDE is obtained from the 2PI CGEA, while the BSEs are obtained from the SDE by differentiation with respect to the external field. The important point is that, because the 2PI CGEA is invariant under external gauge transformations, both the full propagator and response functions obtained this way are external gauge covariant. As we will show below, the EGI property of the 2PI CGEA implies that $G$ and $\Pi_{(n+2)}$, as obtained from $\Gamma[G,A]$, satisfy the WT hierarchy. This provides the required link between current conservation in nonlinear response and the external gauge invariance of the 2PI CGEA, and also a powerful and systematic way of studying nonlinear response in strongly interacting systems coupled to ideal reservoirs.

To see the connection between EGI and the WT hierarchy, recall that external gauge invariance of the effective action means $G[A']=UG[A]U^{-1}$. Inserting the expansion in powers of the external field given in Eq. (\ref{expans}) into both sides of this identity we get
\begin{widetext}
\begin{equation}
\begin{split}
& G[0] + i A'_\mu \Pi_3^\mu + \frac{i^2}{2}A'_\mu A'_\nu \Pi_4^{\mu \nu} + \ldots \\
&= \sum_{n=0} \frac{i^n}{n!} \int d1\ldots dn A'_{\mu_1}(1)\ldots 
A'_{\mu_n}(n) \Pi_{(n+2)}^{\mu_1\ldots \mu_n}(1,\ldots,n; X,Y) \\
&= UG[0]U^{-1} + iA_\mu U\Pi_3^\mu U^{-1} + \frac{i^2}{2}A_\mu A_\nu U\Pi_4^{\mu \nu} U^{-1} + \ldots \\
&= \sum_{n=0} \frac{i^n}{n!} \int d1\ldots dn A_{\mu_1}(1)\ldots 
A_{\mu_n}(n) [U(X)\Pi_{(n+2)}^{\mu_1\ldots \mu_n}(1,\ldots,n; X,Y)U^{-1}(Y)] ~.
\end{split}
\label{egiward}
\end{equation}
\end{widetext}
Note that Eq. (\ref{egiward}) explicitly shows that the response functions $\Pi_{(n+2)}$ are external gauge covariant. 
In particular, for an infinitesimal external gauge transformation $U(X)\approx 1 + ie\varphi(X)$ the transformed external field is 
\begin{equation}
A'_\mu(X) = A_\mu(X) + \partial_\mu \varphi(X) \qquad ,
\end{equation}
so Eq. (\ref{egiward}) becomes
\begin{widetext}
\begin{equation}
\begin{split}
& \sum_{n=0} \frac{i^n}{n!} \int [dn]\prod_{i=1}^n \{A_{\mu_i}(i)+\partial_{\mu_i}\varphi(i)\}\Pi_{(n+2)}^{\mu_1\ldots \mu_n} =\\
& \sum_{n=0} \frac{i^n}{n!} \int [dn]\prod_{i=1}^n A_{\mu_i}(i)\{\Pi_{(n+2)}^{\mu_1\ldots \mu_n}+ie[\varphi(X)\Pi_{(n+2)}^{\mu_1\ldots \mu_n}-\Pi_{(n+2)}^{\mu_1\ldots \mu_n}\varphi(Y)]\}
\end{split}
\label{produc}
\end{equation}
\end{widetext}
where we have suppressed the arguments of $\Pi_{(n+2)}^{\mu_1\ldots \mu_n}(1\ldots n;X,Y)$ and defined $[dn]=d1\ldots dn$ for brevity. Comparing terms of the same order in $A_\mu$ on both sides of this expression we get
\begin{equation}
\begin{split}
& \Pi_2(X,Y)+ i \int d1 [\partial_\mu^1 \varphi(1)]\Pi_3^\mu(X,Y;1) =\\
& \Pi_2(X,Y) + ie [\varphi(X)\Pi_2(X,Y)-\Pi_2(X,Y)\varphi(Y)]
\end{split}
\label{zerotha}
\end{equation}
for the zeroth order term, and
\begin{equation}
\begin{split}
& \int d1 d2 [\partial_\nu^2 \varphi(2)]\Pi_4^{\mu\nu}(X,Y;1,2)A_\mu(1) =\\
& \int d1 e[\varphi(X)\Pi_3^\mu(X,Y;1)-\Pi_3^\mu(X,Y;1)\varphi(Y)] A_\mu(1)
\end{split}
\label{linear}
\end{equation}
for the linear term. Higher order terms have a similar structure but can become quite involved. The main point to emphasize is that the EGI of the 2PI CGEA implies relationships among the response functions, shown explicitly in Eq. (\ref{produc}).

Assuming that $\varphi$ vanishes at infinity, we can integrate by parts the second term in Eq. (\ref{zerotha}) (zeroth order in $A_\mu$) to obtain 
\begin{equation}
\begin{split}
& i\int d1 \varphi(1)\partial_\mu^1 \Pi_3^{\mu}(X,Y;1) =\\
& -i e[\varphi(X)\Pi_2(X,Y)-\Pi_2(X,Y)\varphi(Y)]
\end{split}
\end{equation}
which implies
\begin{equation}
\partial_\mu^z \Pi_3^\mu(X,Y;z)=-e\Pi_2(X,Y)[\delta(X-z)-\delta(Y-z)] 
\end{equation}
or, in more detail, 
\begin{equation}
\begin{split}
& \partial_\mu^z <T_c j^\mu(z)\psi(X)\bar\psi(Y)> = \\
& e<T_c \psi(X)\bar\psi(Y)> [\delta(Y-z)-\delta(X-z)] \qquad .
\end{split}
\label{zeroward}
\end{equation}
This is precisely the identity corresponding to $n=1$ in the WT hierarchy given by Eq. (\ref{jer}), that is, Eq. (\ref{ejemplo1}). We see that, even at zeroth order in the external field, the relation between the three-point vertex and the two-point function, Eq. (\ref{zeroward}), is satisfied due to the EGI of the 2PI EA. 

Similarly, for the linear term, Eq. (\ref{linear}) implies the following identity
\begin{equation}
\begin{split}
&-\int d2 \partial_\nu^2 \Pi_4^{\mu\nu}(X,Y;1,2)\varphi(2) = \\
& e[\varphi(X)\Pi_3^\mu(X,Y;1)-\Pi_3^\mu(X,Y;1)\varphi(Y)]
\end{split}
\end{equation}
which leads by the same calculation as in the zeroth order case  to a WT identity between the three- and four-point response functions (equivalent to Eq. (\ref{jer}) for $n=2$):
\begin{equation}
\begin{split}
& \partial_\nu^z \Pi_4^{\mu\nu}(X,Y;1,z) = \\
& e[\delta(Y-z)-\delta(X-z)]\Pi_3^\mu(X,Y;1) \qquad .
\end{split}
\end{equation}

It is clear that this procedure could be continued to higher order terms in $A_\mu$, thus generating higher order WT identities. We note that the hierarchy obtained for the response functions $\Pi_{(n+2)}$ is completely equivalent to that involving $\Lambda_{(n)}$, given in Eq. (\ref{jer}), as expected since, ultimately, they both enforce current conservation. As it can be seen from their definitions, Eqs. (\ref{lam}) and (\ref{pidef}), the relation  between vertex and response functions can be compactly written as   
\begin{widetext}
\begin{equation}
\Pi^{\mu_1\ldots \mu_n}_{(n+2)}(X,Y;z_1\ldots z_n) = 
(-e)^{(n-1)}\prod_{\alpha=2}^n \lim_{z'_\alpha \rightarrow z_{\alpha}} D^{\mu_\alpha}(z'_{\alpha},z_{\alpha}) \Lambda_{(n)}^{\mu_1}(Xz_2\ldots z_n,Yz'_2\ldots z'_n;z_1)
\label{pilam}
\end{equation}
\end{widetext}
for $n \geq 2$, and 
\begin{equation}
\begin{split}
& \Pi_{(3)}^\mu(X,Y;z)=\Lambda_{(1)}^\mu (X,Y;z) \qquad \textrm{for}\qquad n=1, \\
& \Pi_{(2)}(X,Y)=iG[0](X,Y) \qquad \textrm{for}\qquad n=0 ,
\end{split}
\end{equation}
where the operator $D^{\mu_\alpha}(z'_{\alpha},z_{\alpha})$ is defined in Eq. (\ref{curr}). 
Defining the operator (with $n\geq 2$ and $2 \leq \alpha \leq n$)
\begin{equation}
F_{n}^{\mu_2\ldots \mu_n}(\{z'_\alpha,z_\alpha\})=(-e)^{(n-1)} \prod_{\alpha=2}^n \lim_{z'_\alpha \rightarrow z_{\alpha}} D^{\mu_\alpha}(z'_{\alpha},z_{\alpha})
\end{equation}
that appears in Eq. (\ref{pilam}) acting on $\Lambda_{(n)}^{\mu_1}$, noting that this operator does not depend on $z_1$ (the ``external'' coordinate in $\Lambda_{(n)}$), and using the property of the Green's function $G_n$ 
\begin{equation}
\begin{split}
& F_{n}^{\mu_2\ldots \mu_n}(\{z'_\alpha,z_\alpha\}) G_n(1\ldots z_\alpha \ldots n,1'\ldots z'_\alpha \ldots n') = \\
& F_{n} <T_c \psi(1)\ldots \psi(z_\alpha) \ldots \psi(n)\bar{\psi}(n')\ldots \bar{\psi}(z'_\alpha) \ldots \bar{\psi}(1')> \\
&= <T_c j^{\mu_2}(z_2)\ldots j^{\mu_n}(z_n)\psi(1)\bar{\psi}(1')> =\\
& \Pi_{(n+2)}^{\mu_2\ldots\mu_n}(1,1';z_2\ldots z_n)
\end{split}
\end{equation}
the equivalence between WT hierarchies can be proven using Eq. (\ref{pilam}) (the calculation is tedious but rather straightforward). 

Therefore, the external gauge invariance of the 2PI CGEA enforces the WT hierarchy necessary for current conservation beyond the expectation value level. By expanding the full propagator (solution to the SDE) in powers of the external field, we can calculate current-conserving response functions as solutions to the BSEs obtained from the SDE. The key point is that the EGI of the 2PI CGEA implies the covariance of the full propagator and the response vertex functions, and this results in the WT relationships among them. We emphasize that such a procedure relies on an expansion of $G[A]$ in powers of the external field $A$, so it is valid only for weak external fields. 

We end up this section by indicating how to calculate the current induced by the external classical field. The average current which evolves by the action of the external field is given by Eq. (\ref{curr}), which can be re-expressed in terms of the lesser Green function $G_{+-}$:
\begin{equation}
\begin{split}
<j^\mu(z)> &= -e \lim_{z' \rightarrow z} D^\mu(z,z')[A]<\psi^\dagger(z')\psi(z)> =\\
& ie \lim_{z' \rightarrow z} D^\mu(z,z')[A] G_{+-}(z,z')
\end{split}
\end{equation}
where $D^\mu[A]$ is given in Eq. (\ref{dop}) and one should recall that $G_{+-}$ must be calculated in the presence of the external field\cite{morg}. 
The lesser Green function appearing in the expression for the current can be calculated directly from the expansion given in Eq. (\ref{expans}).
In this way, the induced average current can be systematically calculated to arbitrary order in the external classical field. The approach based on the 2PI EA and its loop truncation, adopted here, guarantees that the WT hierarchy is fulfilled. 

\section{Approximate 2PI effective actions and current conservation}
\label{approxsec}

As shown in Sec. \ref{ctpea}, the virtue of the 2PI CGEA method is that it provides a systematic way of encompassing interacting fields and current-conserving nonlinear response theory in a unified way. In particular, truncations to the loop expansion of the 2PI CGEA are, to arbitrary order, external gauge invariant. Therefore, any such approximation will preserve the WT hierarchy. In this section, we give a concrete example of an {\it ad hoc} approximation to the 2PI CGEA, i.e. not obtained from a truncation in the loop expansion, and analyze the consequences for current conservation. 

To be specific, we consider the following bare interaction vertex in $S_{int}$ as appearing in Eq. (\ref{sint}):
\begin{equation}
U_{ABCD}=c_{abcd}\bar{\delta}_{t,r}\bar{\delta}_\sigma \tilde{U}
\label{uloc}
\end{equation}
where 
\begin{equation}
c_{abcd} = 
\begin{cases}
1 \qquad ~~\textrm{if} ~ a,b,c,d = + \\
-1 \qquad \textrm{if} ~ a,b,c,d = - \\
0 \qquad ~~\textrm{otherwise} 
\end{cases}
\end{equation}
is a CTP tensor. In Eq. (\ref{uloc}) we have grouped delta functions as follows
\begin{equation}
\begin{split}
\bar{\delta}_{t,r} &= \delta(t_A-t_B)\delta(t_A-t_C)\delta(t_A-t_D) \times \\
&~~~~ \delta(r_A-r_B)\delta(r_A-r_C)\delta(r_A-r_D) \\
&= \delta(X_A-X_B)\delta(X_A-X_C)\delta(X_A-X_D)
\end{split}
\end{equation}
and ($\sigma$ stands for spin projection)
\begin{equation}
\bar{\delta}_\sigma = \delta_{\sigma_A,\sigma_B}\delta_{\sigma_C,\sigma_D} \qquad .
\end{equation}
This vertex describes a ``Hubbard-like'' local interaction $n_{\uparrow}(r)n_{\downarrow}(r)$, where $n_{\sigma}(r)$ is the electron number operator at position $r$.

It will be convenient for what follows to write the quantum correction to the 2PI EA as
\begin{equation}
\Gamma_2[G]=\sum_{i=1}^\infty \Gamma_2^{(i)}
\label{gamu}
\end{equation}
where $\Gamma_2^{(i)}$ is of order $\tilde{U}^i$. We note that, due to the fact that fermion mean fields vanish, an expansion of $\Gamma_2$ in powers of $U$ is equivalent to a loop expansion (this is no longer true when symmetry breaking occurs, see e.g. Ref. \onlinecite{libro}). The $O(U)$ diagram corresponds to the double-bubble, which is two-loop, while the $O(U^2)$ one corresponds to the basketball, which is three-loop. Both diagrams are given in Eq. (\ref{gam2}). Since the CTP self-energy is obtained by functionally differentiating $\Gamma_2$ with respect to the full propagator, the contributions to the self-energy can also be classified according to Eq. (\ref{gamu}). According to Eq. (\ref{uloc}), for $i=1$ we have
\begin{widetext}
\begin{equation}
\begin{split}
\Sigma_{AB}^{(1)} =& \Sigma^{(1)}_{ab}(X_A,X_B) = \frac{1}{4}U_{ABCD}G_{CD} \\
=& \frac{1}{4} \tilde{U} c_{abcd} \sum_{\sigma_C,\sigma_D}\int dX_C dX_D \delta(X_A-X_B)\delta(X_A-X_C)\delta(X_A-X_D) \times \\
& \delta_{\sigma_A\sigma_B}\delta_{\sigma_C\sigma_D} G_{cd}(X_C,X_D) \\
=& \frac{1}{2} \tilde{U} c_{abcd} \delta(X_A-X_B) \delta_{\sigma_A\sigma_B} G_{cd}(X_A,X_B) \qquad .
\end{split}
\end{equation}
\end{widetext}
We note that the exchange part of $\Sigma_{AB}$ is absent because the interaction occurs between particles with opposite spin projection. 

From this equation, the $(+,-)$ components of the CTP self-energy are immediately obtained:
\begin{equation}
\begin{split}
\Sigma_{++}^{(1)}(X_A,X_B) &= \tilde{U} \delta(X_A-X_B) \delta_{\sigma_A\sigma_B} G_{++}(X_A,X_B) \\
\Sigma_{--}^{(1)}(X_A,X_B) &= - \tilde{U} \delta(X_A-X_B) \delta_{\sigma_A\sigma_B} G_{--}(X_A,X_B) \\
\Sigma_{+-}^{(1)} &= \Sigma_{-+}^{(1)} = 0
\end{split}
\label{sig1}
\end{equation}
where the last line follows from the definition of the CTP tensor $c_{abcd}$, and shows the well-known fact that the quasiparticle life-time is infinite in the Hartree approximation. 

One can proceed similarly for $i=2$ to obtain (we omit the calculation since it is completely analogous to the $i=1$ case) 
\begin{equation}
\begin{split}
\Sigma_{++}^{(2)}(X_A,X_B) &= -i\tilde{U}^2 \delta(X_A-X_B) \delta_{\sigma_A\sigma_B} [G_{++}(X_A,X_B)]^3 \\
\Sigma_{--}^{(2)}(X_A,X_B) &= -i\tilde{U}^2 \delta(X_A-X_B) \delta_{\sigma_A\sigma_B} [G_{--}(X_A,X_B)]^3 \\
\Sigma_{+-}^{(2)}(X_A,X_B) &= i\tilde{U}^2 \delta(X_A-X_B) \delta_{\sigma_A\sigma_B} [G_{+-}(X_A,X_B)]^3 \\
\Sigma_{-+}^{(2)}(X_A,X_B) &= i\tilde{U}^2 \delta(X_A-X_B) \delta_{\sigma_A\sigma_B} [G_{-+}(X_A,X_B)]^3 \\
\end{split}
\label{sig2}
\end{equation}
showing that this approximation includes fluctuation damping since the lesser and greater components of the self-energy are nonzero. The full CTP self-energy to order $U^2$ is then given by $\Sigma^{(1)}+\Sigma^{(2)}$.

As we have shown in previous sections, the truncation of the loop ($U$) expansion of $\Gamma_2[G]$ does not violate the external gauge invariance of the full 2PI EA, therefore providing approximate propagator and vertex functions that satisfy the WT hierarchy. We will now discuss, in the context of the 2PI EA description presented in this work, an {\it ad hoc} approximation known in non-equilibrium perturbation theory applied to transport through quantum dots (see Refs. \onlinecite{pump,hersh} and references therein), and show why it does not preserve current conservation. 

In the language of the 2PI EA formalism, the approximation involves two separate steps. In the first one, the quantum correction $\Gamma_2$ is approximated by the two-loop ($O(U)$) contribution, $\Gamma_2^{(1)}$. The $O(U^2)$ self-energy is calculated self-consistently either from\cite{hersh} the {\it Hartree} $G_{+-}$ (lesser correlation) or by requiring\cite{yey,pump} that the occupation of the central region evaluated with a renormalized $\Sigma^{(1)}$ (but not the Hartree one) equals the one calculated (in the next step) from the propagator dressed with $\Sigma^{(2)}$. In the second step, $\Sigma^{(2)}$ is calculated from Eq. (\ref{sig2}) but with the full propagator $G_{AB}$ replaced by that calculated from $\Sigma^{(1)}$ in the first step. 
In simpler terms, in this approximation the internal lines in $\Sigma^{(2)}$ are not the full propagator, as they are in the 2PI EA approach, but are either the Hartree or a similarly calculated propagator. It is easy to recognize that such approach can not, as it is, produce current conserving results in general, since it breaks the variational procedure by which the full propagator is obtained from a single (truncated) functional.

The difference between the approximations to the 2PI EA based on a loop truncation and the {\it ad hoc} one can be best appreciated by comparing their SD equations. For the truncation to three loops we have
\begin{equation}
G_{AB}^{-1} = D_{AB}^{-1} + i \Sigma_{AB}^{(1)}[G] + i \Sigma_{AB}^{(2)}[G]
\label{loop3}
\end{equation}
where we have made explicit that $\Sigma^{(1,2)}$ are functionals of the full CTP propagator $G_{AB}$, given by Eqs. (\ref{sig1}) and (\ref{sig2}) respectively. For the {\it ad hoc} approximation we have instead the following system of self-consistent equations:
\begin{equation}
\begin{split}
& g^{-1}_{AB} = D^{-1}_{AB}+i\Sigma^{(1)}_{AB}[g]+K \\
& \tilde{g}_{AB}^{-1} =D^{-1}_{AB} +i\Sigma^{(1)}_{AB}[g]+i\Sigma^{(2)}_{AB}[g] \\
& \tilde{g}_{+-}(r,t;r,t) = g_{+-}(r,t;r,t)
\end{split}
\label{adhoc}
\end{equation}
where $\tilde{g}$ is the $O(U^2)$ propagator in this approximation scheme, and $K$ is a source added in the calculation of $g$ to enforce the condition expressed in the third line. This condition corresponds to requiring that the level occupation of the central region calculated from $g$ and from $\tilde{g}$ be equal. Note that if $K=0$ then the $O(U)$ propagator, $g$, becomes the usual Hartree propagator. The equation of motion for $\tilde{g}$ can be cast in a more familiar form\cite{pump,kel} using the properties of CTP propagators described in Sec. \ref{wtegisec}. The result is ($1=(r_1,t_1)$)
\begin{widetext}
\begin{equation}
\begin{split}
& \tilde{g}_{+-}(1,2) = \int d3 d4~ \tilde{g}_r(1,3) \{ \Sigma_{\phi,+-}(3,4) + \Sigma^{(2)}_{+-}[g](3,4)\} \tilde{g}_a (4,2) \\
& \tilde{g}_{-+}(1,2) = \int d3 d4 ~\tilde{g}_r(1,3) \{ \Sigma_{\phi,-+}(3,4) + \Sigma^{(2)}_{-+}[g](3,4)\} \tilde{g}_a (4,2) \\
& \tilde{g}_{r}(1,2) = g_r(1,2) +\int d3 d4 ~ g_r(1,3) \Sigma^{(2)}_{r}[g](3,4) \tilde{g}_r(4,2) \qquad ,
\end{split}
\label{gfam}
\end{equation}
\end{widetext}
where the retarded $O(U^2)$ self-energy is
\begin{equation}
\Sigma^{(2)}_r (1,2) = \theta(1,2)[\Sigma^{(2)}_{+-}(1,2)-\Sigma^{(2)}_{+-}(1,2)] 
\end{equation}
and $\Sigma_{\phi}$ is given in Eq. (\ref{sigmaphi}). Using that the reservoirs are in equilibrium and non-interacting, and that the coupling matrix does not mix CTP branches, i.e. $T_{AB}=\textrm{diag}(T,T)$, the components of $\Sigma_{\phi}$ in energy-momentum space can be written as 
\begin{equation}
\begin{split}
& \Sigma_{\phi,+-}(p) = |T|^2 B_{+-}(p) = \sum_j 2\pi i |T|^2 n(p)\delta(p_0-p_j^2/2) \\
& \Sigma_{\phi,-+}(p) = |T|^2 B_{-+}(p) \\
&= -\sum_j 2\pi i |T|^2 [1-n(p)]\delta(p_0-p_j^2/2)
\end{split}
\end{equation}
where $n(p)$ is Fermi-Dirac function, the momentum index $j$ denote single-particle energy levels, and we have assumed that $T(p)=T$, independent of $p$. 
We note that the term describing the evolution of initial correlations is absent from Eq. (\ref{gfam}), since we are assuming that the system is in a non-equilibrium stationary (or periodic) state mantained by the external classical field. 

The equation of motion for the full propagator $G$ has the same structure as the one for $\tilde{g}$, with the following replacements 
\begin{equation}
\begin{split}
& \tilde{g}_{+-},\tilde{g}_{-+},\tilde{g}_r \rightarrow G_{+-}, G_{-+}, G_r \\
& \Sigma^{(2)}_{+-}[g], \Sigma^{(2)}_{-+}[g], \Sigma^{(2)}_r[g] \rightarrow \Sigma_{+-}[G],\Sigma_{-+}[G],\Sigma_r[G] \\
& g_r \rightarrow D_r  \qquad .
\end{split}
\end{equation}

We will now consider the expectation value of the electronic current in the {\it ad hoc} approximation discussed so far. As emphasized in Sec. \ref{wtintro}, the WT hierarchy given in Eq. (\ref{jer}) is a direct consequence of the equations of motion satisfied by the {\it exact} propagators. This can be shown quite simply by noting that the $(n+1)-$point vertex function $\Lambda^\mu_{(n)}$, defined in Eq. (\ref{lam}), can be rewritten in terms of the operator $D^\mu(z,z')$, defined in Eq. (\ref{dop}): 
\begin{equation}
\begin{split}
& \Lambda_{(n)}^\mu(1\ldots n,1'\ldots n';z)= \\
& -e i^{(n+1)}\lim_{z'\rightarrow z}D^\mu(z,z')G_{(n+1)}(1\ldots n z,1'\ldots n' z') \qquad .
\end{split}
\end{equation}
The divergence of $\Lambda^\mu_{(n)}$ will therefore be given by the divergence of the operator $D^\mu(z,z')$, which follows immedeately from its definition. The result is
\begin{widetext}
\begin{equation}
\begin{split}
& \partial_\mu \Lambda_{(n)}^\mu(1\ldots n,1'\ldots n';z)=\\
& i^n e \lim_{z'\rightarrow z} \big\{ [i(\partial_{t_z} + \partial_{t_{z'}})+\frac{1}{2}(\nabla^2_z-\nabla^2_{z'})-A_i(z)+A_i(z')]G_{(n+1)}(1\ldots z;1'\ldots z')\big\} =\\
& i^n e \lim_{z'\rightarrow z} \zeta(z,z')G_{(n+1)}(1\ldots z;1'\ldots z') \qquad .
\end{split}
\label{dlam}
\end{equation}
\end{widetext}
This equation shows that the divergence of the expectation value of the four-current $j^\mu(z)$, that corresponds to $n=0$, is directly given by the action of the differential operator $\zeta(z,z')$ acting on $G_{(1)}(z,z')=G(z,z')$. Using Heisenberg's equation of motion for the field operators $\psi,\psi^\dagger$ it is easy to show\cite{kb,rivier} that the exact propagator satisfy two differential equations 
\begin{widetext}
\begin{equation}
\begin{split}
[i\partial_{t_z}+\nabla^2_z - A_i(z)]G(z,z') &= \delta(z,z')-i\int d1~ U(z,1) G_2(z1,z'1^+) \\
[-i\partial_{t_{z'}}+\nabla^2_{z'} - A_i(z')]G(z,z') &= \delta(z,z')-i\int d1~ U(1,z') G_2(z1^-,z'1)
\end{split}
\label{diffeqs}
\end{equation}
\end{widetext}
that, when substracted, result in 
\begin{equation}
\lim_{z'\rightarrow z} \zeta(z,z')G(1\ldots z;1'\ldots z') = 0 \qquad ,
\label{lgzero}
\end{equation}
whereby $\partial_\mu <j^\mu(z)>=0$ in the exact theory. 

In an approximate theory, the quantity $-iUG_2$ appearing in Eq. (\ref{diffeqs}) is replaced by the product $\Sigma G$, and so the conservation of current at the expectation value is not immedeately guaranteed. It is well-known\cite{baym,dedom,libro,selfcons} that $\Phi-$derivable approximations are such that $\partial_\mu <j^\mu(z)>=0$. As we have shown in previous sections, the approximations to the 2PI EA based on truncations of the loop expansion guarantee not only this condition but the entire WT hierarchy (order by order in the expansion of $G$ with respect to the external driving field). 

The approximation defined by Eq. (\ref{adhoc}) can not be derived from a loop-truncated effective action, and hence Eq. (\ref{lgzero}) is not expected to hold. To end up this section, we will briefly show how large is the divergence of the current expectation value in the {\it ad hoc} approximation.
From Eq. (\ref{diffeqs}) we have that, for an approximation based on a truncation of the loop expansion of the 2PI EA, the following identity holds: 
\begin{equation}
\begin{split}
&\lim_{z'\rightarrow z} \zeta(z,z') G(z,z') = \\
& \int d1~ \{ G(z,1)\Sigma^{(2)}[G](1,z)-\Sigma^{(2)}[G](z,1)G(1,z) \} = 0
\end{split}
\label{exactdiv}
\end{equation}
where we note that the Hartree part of the self-energy, $\Sigma^{(1)}[G]$, can be included in the left hand side of Eq. (\ref{diffeqs}) as an extra single-particle term, and, being local, its difference vanishes when $z'\rightarrow z$. Therefore, in Eq. (\ref{exactdiv}) we only need to consider $\Sigma^{(2)}$. Putting $G = \tilde{g} + \tilde{G}$ and $\Sigma^{(2)}[G]=\Sigma^{(2)}[g]+\Xi$ in Eq. (\ref{exactdiv}), and neglecting terms proportional to $\tilde{G}$ and $\Xi$, we immedeately obtain 
\begin{equation}
\begin{split}
& \partial_\mu <\tilde{j}^\mu(z)> =\\
& e\int d1~ \{ \tilde{g}(z,1)\Sigma^{(2)}[g](1,z)-\Sigma^{(2)}[g](z,1)\tilde{g}(1,z) \} 
\end{split}
\end{equation}
where $\tilde{j}$ denotes the current calculated in the {\it ad hoc} approximation. Since $\tilde{g},\Sigma^{(2)}[g] \sim O(U^2)$, we conclude that the violation of current conservation in the mean is $O(U^4)$ or higher. 

To conclude, we note that the combination of Eqs. (\ref{dlam}) and (\ref{exactdiv}) constitute a useful way of checking that a given approximation is conserving in the mean, and may provide some insight in the search of conserving approximations not based on loop-truncations. As we have shown, it also provides a way of calculating an upper bound to the violation of mean current conservation.

\section{Conclusions}
\label{concsec}

In this work, we have determined the basic requirements that an approximation to a non-equilibrium many-body problem in an open and driven fermionic system must satisfy in order to achieve current conservation beyond the expectation value level. One of the most important results of this work is the close relation found between nonlinear response theory and the Ward-Takahashi hierarchy, necessary for current conservation. This connection, although already known\cite{dah,bon,vel,myo,kr,baym,kb}, was clearly displayed by using the closed-time path two-particle irreducible coarse-grained effective action to describe the central electrons coupled to ideal reservoirs. 

We have shown that the gauge invariance of the 2PI effective action with respect to transformations of the external classical field driving the system automatically implies the WT hierarchy among the propagator and vertex functions calculated from the effective action. More importantly for practical calculations, for every approximation to the 2PI effective action that results from truncating its loop expansion, the closed-time path propagator (obtained from the Schwinger-Dyson equation) and the the vertex functions (obtained from the Bethe-Salpeter equations) are such that the WT hierarchy holds. 

Using a simple example, we have also discussed, in the context of the 2PI effective action formalism, the relation between {\it ad hoc} approximations (not obtained from a truncation of the loop expansion) and current conservation. Using a general expression for the divergence of the mean current, we have shown that in the {\it ad hoc} approximation considered current conservation at the expectation value level is violated at $O(U^4)$ or even at higher order. 

In summary, we hope to have shown that closed-time path 2PI effective action techniques are a powerful and systematic method to study the nonlinear response in strongly correlated open systems in weak external fields, in a current-conserving way. We also hope to have shown the necessity, within the EA approach, of using loop-truncated approximations to the 2PI EA when going beyond the linear response regime. Our results may be of use in the theoretical study of quantum transport through interacting electronic pumping devices, which are nowadays receiving much attention.

\begin{acknowledgments}
We acknowledge Liliana Arrachea, Carlos Na\'on, Alfredo Levy Yeyati and Sergio Ulloa for useful discussions. This work has been supported in part by ANPCyT, CONICET and UBA (Argentina).
\end{acknowledgments}

\end{document}